\documentclass[aps,prl,superscriptaddress,twocolumn,amssymb,amsmath]{revtex4-1}
\usepackage{graphicx}
\usepackage{dcolumn}
\usepackage{bm}

\newcommand{\eqn}[1]{\begin{equation}#1\end{equation}}

\begin{document}

\preprint{APS/123-QED}
\title{Realization of a quantum random generator certified with the Kochen-Specker theorem}

\author{Anatoly Kulikov}
\email{a.kulikov@uq.edu.au}
\affiliation{ARC Centre of Excellence for Engineered Quantum Systems, Queensland 4072, Australia}
\affiliation{School of Mathematics and Physics, University of Queensland, St Lucia, Queensland 4072, Australia}

\author{Markus Jerger}
\affiliation{ARC Centre of Excellence for Engineered Quantum Systems, Queensland 4072, Australia}
\affiliation{School of Mathematics and Physics, University of Queensland, St Lucia, Queensland 4072, Australia}

\author{Anton Poto\v{c}nik}
\affiliation{Department of Physics, ETH Zurich, CH-8093 Zurich, Switzerland}

\author{Andreas Wallraff}
\affiliation{Department of Physics, ETH Zurich, CH-8093 Zurich, Switzerland}

\author{Arkady Fedorov}
\affiliation{ARC Centre of Excellence for Engineered Quantum Systems, Queensland 4072, Australia}
\affiliation{School of Mathematics and Physics, University of Queensland, St Lucia, Queensland 4072, Australia}

\date{\today}

\begin{abstract}
Random numbers are required for a variety of applications from secure communications to Monte-Carlo simulation. Yet randomness is an asymptotic property and no output string generated by a physical device can be strictly proven to be random. We report an experimental realization of a quantum random number generator (QRNG) with randomness certified by quantum contextuality and the Kochen-Specker theorem. The certification is not performed in a device-independent way but through a rigorous theoretical proof of {\it each} outcome being value-indefinite even in the presence of experimental imperfections. The analysis of the generated data confirms the incomputable nature of our QRNG.
\end{abstract}

\pacs{Valid PACS appear here}
\maketitle


While we can consider a mathematical abstraction of a true random number generator and examine its properties, in the physical world we are confined to performing finite statistical tests on the output strings. By applying sets of such tests (like NIST~\cite{Rukhin2010} or diehard~\cite{Marsaglia1996}) we can verify with arbitrarily high probability that the generator is NOT random (if it has failed at least one test), but cannot prove its randomness in the opposite case. As an example, one may construct a pseudo-random number generator which passes all above-mentioned tests while the produced sequence is deterministic and even computable \cite{Matsumoto1998}.
The impossibility of a rigorous proof of randomness for a finite string generated by a physical device motivates the consideration of more fundamental arguments to support a RNG's randomness. From this point of view, no classical RNG may be truly random as it is deterministic by the laws of classical mechanics, and may in principle be predicted. A natural foundation to build a RNG would be quantum theory, as it is intrinsically random.

However, although quantum mechanics obeys probabilistic rules, the possibility of separating intrinsic randomness from apparent randomness arising from  a lack of control or from experimental noise is still under debate \cite{dhara2014}. Moreover, while quantum mechanics for a two-level system is described by the same intrinsically-probabilistic measurement rules, one may not strictly prove value-indefiniteness, and hence indeterminism, of its results \cite{Calude2008}. 

These considerations led to the next advance in quantum number generation: the protocols certified by violation of certain Bell-type inequalities \cite{Pironio2010,Fehr2013,Herrero-Collantes2017}. More specifically, through violation of the CHSH inequality one may certify that the observed outputs are not entirely predetermined and write a lower bound on the generating process entropy. Unfortunately, this approach does not allow one to close the gap between this lower bound and true randomness. In addition, the Bell-type certification schemes can be regarded as random expanders rather than generators due to the requirement of ``a small private random seed" to operate \cite{Abbott2012,Pironio2010, Um2013}. Finally, the random number generators certified by  Bell inequalities utilize no-signaling assumption and is, therefore, inherently a non-local device which is challenging to use for practical applications.

To address this problem, a different approach to QRNG certification based on the Kochen-Specker theorem and contextual measurements has recently been proposed \cite{Abbott2012}. It does not allow certification of the data in a device-independent fashion like the CHSH inequality, but yields a rigorous theoretical proof of measurements outcomes being value-indefinite even in the presence of experimental imperfections. 
In this Letter we experimentally realize a random number generator certified by the Kochen-Specker theorem. We use the circuit quantum electrodynamics (QED) as a platform for the physical realization of a QRNG. A superconducting qutrit has been used recently to demonstrate quantum contextuality, the resource underlying the operation of the QRNG~\cite{Jerger2016a}. Utilizing high controllability and fast repetition rates for circuit QED devices we reach a bit rate two orders of magnitude higher than previously reported certified random number generators \cite{Pironio2010, Herrero-Collantes2017, Um2013}.


\begin{figure}[t]
	\centering
	\includegraphics[width=0.49\textwidth]{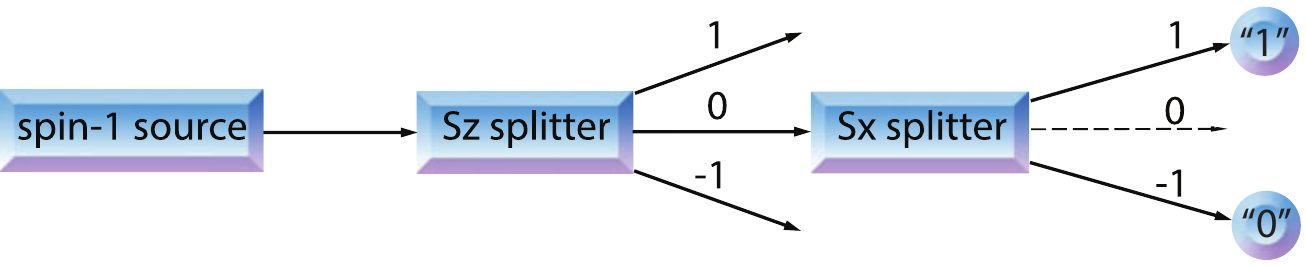}
	\caption{The theoretical protocol of the QRNG certified by the strong Kochen-Specker theorem proposed in Ref.~\cite{Abbott2012}. The protocol is formulated for a spin-1 particle and consists of two sequential measurements. The first measurement is used to initialize the particle in the $S_z = 0$ eigenstate of the spin operator $S_z$. The second measurement is performed in the eigenbasis of the $S_x$ operator with the two outcomes $S_x=\pm 1$ realized randomly as proven by the Kochen-Specher theorem. The outcome $S_x = 0$ is never realized in the ideal case but can be used to monitor the quality of the protocol implementation.
	}
	\label{fig:1}
\end{figure}

\begin{figure*}[ht]
	\begin{center}
			\includegraphics[width=0.85\textwidth]{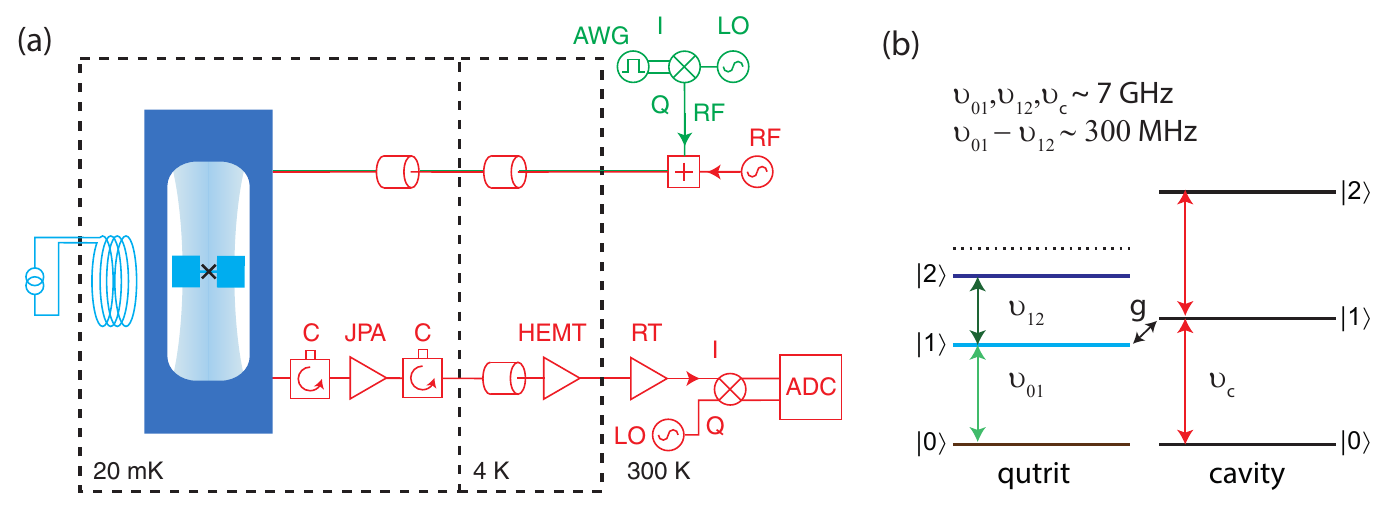} 
			\caption{
		(a) Simplified diagram of the measurement setup. A transmon type multi-level quantum system is incorporated into a 3D microwave copper cavity attached to the cold stage of a dilution cryostat. A magnetically tunable Josephson junction (SQUID) is used to control the transition frequency of the qutrit by a superconducting coil attached to the cavity. Amplitude-controlled and phase-controlled microwave pulses are applied to the input port of the cavity by a quadrature IF (IQ) mixer driven by a local oscillator (LO) and sideband modulated by an arbitrary waveform generator (AWG). The measurement signals transmitted through the cavity are amplified by quantum Josephson parmateric amplifier (JPA), by a high-electron-mobility transistor (HEMT) amplifier at 4~K and a chain of room temperature (RT) amplifiers. The sample at 20~mK is isolated from the higher temperature stages by three circulators (C) in series. The amplified transmission signal is down-converted to an intermediate frequency of 25~MHz in an IQ mixer driven by a dedicated LO, and is digitized by an analog-to-digital converter (ADC) for data analysis.
		(b) The energy level diagram of a qutrit coupled to a microwave cavity. The transition frequencies of the qutrit and cavity are in GHz while the anharmonicity of the qutrit is $\sim 300\,$MHz. When the coupling $g$ between the transmon and the cavity is much smaller than their mutual detuning, the system is in the dispersive regime used for measurement of the qutrit.
		} 
		\label{fig:setup}
	\end{center}
\end{figure*}

To realize the protocol shown in  Fig.~\ref{fig:1} we use a superconducting quantum system, called a transmon, coupled to a microwave cavity. The transmon has a weakly anharmonic multi-level structure~\cite{Koch2007}, and its three lowest energy eigenstates $|0\rangle, |1\rangle$ and $|2\rangle$ are used as the logical states of a qutrit (see Fig.~\ref{fig:setup}). In the dispersive regime, where the cavity resonance frequency is sufficiently detuned from the qutrit transition frequencies, the qutrit-cavity interaction causes cavity frequency shifts dependent on the populations of the energy eigenstates of the transmon~\cite{Koch2007}. These shifts, called dispersive shifts, are extensively used for realizing dispersive readout of superconducting qubits and qutrits by measuring microwave transmission through the cavity (for a specific example of the measurement of a qutrit, see Ref.~\cite{Bianchetti2010,Jerger2016}). 


Manipulations of the qutrit quantum state can be realized with microwave pulses resonant to the $|0\rangle - |1\rangle$ or $|1\rangle - |2\rangle$ transition frequencies and applied to the qutrit through a separate on-chip charge line. In the following we define $R_{\hat n}^{i,i+1}(\phi)$ as rotations of a quantum state of angle $\phi$ about the axis $\hat n$ in the qutrit subspace spanned by $\{|i\rangle, |i+1\rangle\}$. In particular, one can realize the following rotations
\begin{eqnarray}\nonumber
R^{12}_y(\theta) \equiv \left(
	\begin{matrix}
		1 & 0 & 0 \\
		0 &  \cos\theta/2 & \sin\theta/2 \\
		0 & -\sin\theta/2 & \cos\theta/2
	\end{matrix}\right);\\ \label{rot-mat}
	\\ \nonumber
	R^{01}_y(\theta) \equiv \left(
	\begin{matrix}
		\cos\theta/2 & \sin\theta/2 & 0 \\
		-\sin\theta/2 & \cos\theta/2 & 0 \\
		0 & 0 & 1
	\end{matrix}\right).
\end{eqnarray}

\

To reformulate the protocol shown in Fig.~1 in terms of energy eigenstates of the transmon we map the eigenstates of the $S_z$ operator to the states of the qutrit as follows 
\begin{equation}
\{|z,-1\rangle,|z,0\rangle,|z,+1\rangle\} \rightarrow \{|2\rangle,|0\rangle,|1\rangle\}.
\end{equation}
In the eigenbasis of the qutrit the spin-1 operator will take the form
\eqn{
	S_z \equiv  \left(
	\begin{matrix}
		0 & 0 & 0 \\
		0 &  1 & 0 \\
		0 & 0 & -1
	\end{matrix}\right), \mbox{    }
	S_x \equiv \frac{1}{\sqrt{2}} \left(
	\begin{matrix}
		 0& 1 & 1 \\
	1 & 0  & 0 \\
	1 & 0 & 0
	\end{matrix}\right), }
with eigenstates of the operator $S_x$
\eqn{
	|x,-1\rangle =\frac{1}{2} \left(
	\begin{matrix}
	-\sqrt{2} \\
		1  \\
		1 
	\end{matrix}\right), \mbox{    }
	|x,0\rangle =\frac{1}{\sqrt{2}} \left(
	\begin{matrix}
	0 \\
		-1  \\
		1 
	\end{matrix}\right), \mbox{    }
|x,1\rangle =\frac{1}{2} \left(
\begin{matrix}
	\sqrt{2} \\
	1  \\
	1 
\end{matrix}\right).}

For our qutrit encoding the system is initialized in the ground state $|S_z = 0\rangle = |0\rangle$ by cooling down the transmon to the base temperature of a dilution cryostat ($\sim20\,$mK), thus performing the first measurement in the protocol shown in Fig.~\ref{fig:1}. The spurious thermal population of the excited states has been measured to~be~$<1\%$.

Our dispersive readout realizes a projective measurement of the qutrit described by three operators: $\{|0\rangle\langle0|,|1\rangle\langle1|,|2\rangle\langle2|\}$. In order to perform measurement in the eigenbasis of $S_x$ we followed the standard procedure and performed rotations of the state before and after the dispersive measurement. 
More specifically, to measure some arbitrary state $|\psi\rangle$ in the eigenbasis of the $S_x$ operator, we first apply $M^{\dagger} = R^{01}_y(\pi/2)\cdot R^{12}_y(\pi/2)$ to rotate the state of the qutrit $|\psi\rangle$ before the dispersive measurement

\begin{widetext}
\begin{eqnarray}
M^{\dagger}|\psi\rangle &=& R^{01}_y(\pi/2) R^{12}_y(\pi/2)\left(\alpha_{-1} |x,-1\rangle +\alpha_0 |x,0\rangle +\alpha_1 |x,1\rangle \right) \\ \nonumber
&=&\alpha_{-1} |1\rangle +\alpha_0 |2\rangle +\alpha_1 |0\rangle.
\end{eqnarray}
\end{widetext}
During the dispersive measurement the state is projected to one of the energy eigenstates $|i\rangle$ with probabilities described by $|\alpha_{i}|^2$. Then we can apply an additional rotation $M$ to make the full procedure equivalent to the measurement described by $\{|x,-1\rangle\langle x,-1|,|x,0\rangle\langle x,0|,|x,1\rangle\langle x,1|\}$. Note that the last rotation does not change the outcome of the measurement and was not implemented in the actual protocol.
As the system is initialized in $|0\rangle$ state the measurement will produce outcomes $S_x\pm 1$ encoded as $``1"$ and $``0"$ with equal probabilities while $S_x=0$ outcome will ideally never  be realized. If outcomes $S_x=0$ are detected these traces can be discarded and will not affect the randomness of the generated numbers in accordance with the recipe of Ref.~\cite{Abbott2012}.

\begin{figure*}[ht]
	\begin{center}
			\includegraphics[width=0.85\textwidth]{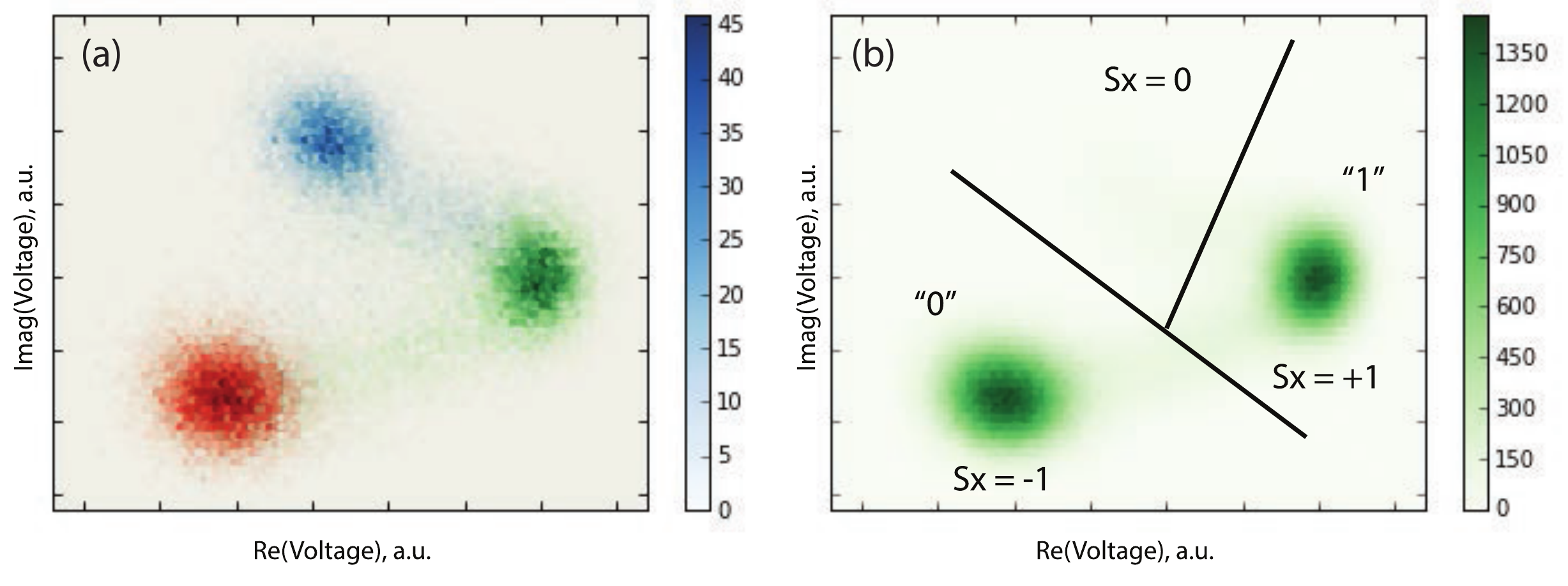} 
			\caption{
		{\bf (a)} Hexbin histogram plot of single-shot three-level readout of different qutrit states. Red: ground qutrit state; Green -- excited state ($|1\rangle$). Blue: second excited state ($|2\rangle$). The intensity of the color represents the number of measurement outcomes falling in each bin.
		{\bf (b)} Hexbin plot of the output of the protocol. Shown are logical encoding of the resulting states and the correspondence to the spin-1 protocol. Note, the $S_x = 0$ state is almost ($<0.1\%$) never realized. The black lines sketch the boundaries of the classification regions.
		} 
		\label{pi2pic}
	\end{center}
\end{figure*}

To distinguish between three different states with high fidelity we use a Josephson parametric amplifier similar to the one described in Ref.~\cite{Eichler2014}. In addition, we set the readout pulse frequency close to the cavity frequency corresponding to the $|1\rangle$ state of the qutrit, which allowed the three possible qutrit states to be well separated on I-Q plane (see Fig.~\ref{pi2pic}(a)). The readout frequency was fine-tuned to maximize the three-level readout fidelity. Using the outlined procedure for initialization and measurement we generated 10 Gbit of raw data at a rate of 50 kbit/s (see Fig.~\ref{pi2pic}(b) for  logical encoding of the resulting states and the correspondence to the spin-1 protocol).

If the qutrit is prepared in the state $|\phi\rangle$ and we perform a quantum measurement described by the projectors $|\psi\rangle\langle\psi|$ then Ref.~\cite{Abbott2012} (improved in Ref.~\cite{Abbott2015}) provides the condition to certify the value-indefiniteness of the outcomes of the measurements:
\begin{equation}\label{condition}
\sqrt{\frac{5}{14}} \leq |\langle\psi|\phi\rangle| \leq \frac{3}{\sqrt{14}}.
\end{equation}
In our protocol we take $\{S_z = 0\}$ state as $|\phi\rangle$ and $\{S_x = \pm 1\}$ as $|\psi_{\pm}\rangle$ (see Fig.~\ref{fig:1}).
If our system were ideally prepared in the ground state and all the experimental imperfections were generated only by errors in the microwave control we could estimate $|\langle\psi_{\pm}|\phi\rangle|$ directly as the square root of the probability to obtain the outcomes $``0"$ and $``1"$. The resulting probabilities to obtain $"0"$ and $``1"$ were measured as $0.536\pm0.004$ and  $0.464\pm0.004$ confirming that the control errors of our setup guarantee value-indefiniteness with high confidence.

In reality the actual states of the system before and after the measurement are not described by pure states. The main contribution to the deviation of the probabilities from the ideal value of $1/2$ is due to relaxation of the qutrit during the dispersive measurement. As it leads to the misinterpretation of the excited state as being the ground state, we measured  greater probability to obtain $``0"$ rather than $``1"$. Another sources of imperfections are thermal excitation of the qutrit ($<1\%$), fidelity of gates ($>99\%$) and misinterpretation of the outcome due to amplifier noise ($0.006\%$). The result of these imperfections may lead to a situation when for some runs the certification condition will not be fulfilled. To provide a confidence low bound for randomness to be certified we conservatively assume that the deviation of the probabilities from the ideal value $1/2$ is only due to the runs where the certification condition (\ref{condition}) is not valid. Thus, we estimate that only $95\%$ of our generated bits are certified random.

As a last step, we address the bias in probabilities of getting $``0"$ and $``1"$ by a standard procedure. For each bit of final data we perform the measurement two times in a row. We encode logical $``0"$ and $``1"$ in the physical events $``01"$ and $``10"$ respectively, which have the same probability to occur, and ignore the two other outcomes. It is straightforward to prove that the properties of QRNG will be preserved: new bits will be certified by value indefiniteness and independent from each other.
This normalization process yields an unbiased sequence with probabilities of $``0"$ and $``1"$ to be $50\%$ each, which is supported by the obtained $50.001\%$ mean frequency of obtaining the $0$ outcome and the standard deviation of $0.1\%$, which is consistent with the bucket size of 999302 raw bits produced. It also increases certification bound: $99.7\%$ of the final bits are certified random: it is sufficient to have one random physical event in the logical sequence to certify the whole sequence to be random.

%
The entropy for the unbiased random numbers obtained from 10 GBit raw data is 7.999999 per byte and is consistent with the ideal value of 8. The data passes all tests in standard NIST and diehard statistical test suites. Moreover, in Ref.~\cite{Abbott2017} the quantum random bits were also analyzed with a test  more directly related to the algorithmic randomness of a sequence (rather than simply statistical properties). Specifically, the raw bits were used to test the primality of all Carmichael numbers smaller than $54\times 10^{7}$ with the  Solovay-Strassen probabilistic algorithm, and the minimum random bits necessary to  confirm compositeness was used as the metric. Ten sequences of raw quantum random bits of length $2^{29}$  were compared with sequences of the same length from three modern pseudo-random generators (Random123, PCG and xoroshilro128+) and a significant advantage was found using the quantum bits. This gives an experimental evidence of the incomputability of the quantum random generator, as predicted by the Kochen-Specker theorem~\cite{Abbott2012}.

In summary, we experimentally demonstrated that the Kochen-Specker certification scheme allows one to eliminate the necessity for input seed random numbers, lifts the non-locality requirements for the certified generator, greatly enhances the rate of generation of certified random numbers, and shows advantage over pseudo-random generators.
The rate of generation of 25 kBit/s of unbiased random bit is limited by the qutrit decay rate ($T_1\sim 5~\mu$s) and may be further increased by using active schemes for initialization of the system in the ground state~\cite{Geerlings2012b,Riste2012}. The certification confidence of $99.7\%$ can be improved by using qutrits with longer relaxation times. 

\begin{acknowledgements}
We would like to thank Cristian S. Calude, Alastair A. Abbott, Michael J. Dinneen and Nan Huang for valuable discussions, comments on the manuscript and clarification on the results of the analysis of the data from Ref.~\cite{Abbott2017}. We thank ETHZ Qudev team for helping us with fabrication of the qubit and parametric amplifier. We also thank Renato Renner for critical reading of our manuscript and providing us with useful feedback.
The authors were supported by the Australian Research Council Centre of Excellence CE110001013. 
\end{acknowledgements}

A.K. conducted the measurements and analysed the data. M.J. built the measurement setup and software. A.P. and A.W. fabricated the parametric amplifier. A.F. supervised the project and fabricated the qubit. A.K. and A.F. wrote the manuscript.



\bibliographystyle{apsrev4-1}
\bibliography{R:/EQUS-SQDLab/RefDB/SQDRefDB}

\end{document}